# EXACT-CT: EXplainable Analysis for Crohn's and Tuberculosis using CT

Shashwat Gupta[1*], Sarthak Gupta[2], Akshan Agrawal[1], Mahim Naaz[3], Rajanikanth Yadav[3], Priyanka Bagade[1*]

[1]CSE, IIT Kanpur, Kanpur, 208016, Uttar Pradesh, India.
[2]Radiology, KMC Mangalore, Mangalore, 575001, Karnataka, India.
[3]Radiology, SGPGI Lucknow, Lucknow, 226014, Uttar Pradesh, India.

*Corresponding author(s). E-mail(s): guptashashwatme@gmail.com;
pbagade@cse.iitk.ac.in;
Contributing authors: akshancs2020@gmail.com;
mahimnaz8858@gmail.com; rajani@sgpgi.ac.in;

**Abstract**

Crohn's disease and intestinal tuberculosis share many overlapping features such as clinical, radiological, endoscopic, and histological features—particularly granulomas, making it challenging to clinically differentiate them. Our research leverages 3D CTE scans, computer vision, and machine learning to improve this differentiation to avoid harmful treatment mismanagement such as unnecessary antituberculous therapy for Crohn's disease or exacerbation of tuberculosis with immunosuppressants. Our study proposes a novel method to identify radiologist-identified biomarkers (VF/SF, necrosis, calcifications, comb sign, pulmonary TB) to enhance accuracy. We demonstrate the effectiveness by using different ML techniques on the features extracted from these biomarkers, computing SHAP on XGBoost for understanding feature importance towards predictions, and comparing against SOTA methods such as pretrained ResNet and CTFoundation.

**Keywords:** Crohn's Disease, Intestinal Tuberculosis, CT Enterography, Explainable AI, Radiological Biomarkers

# 1  Introduction

Crohn's disease (CD) and intestinal tuberculosis (ITB) are chronic granulomatous intestinal disorders whose incidence is rising in South and Southeast Asia [1]. Despite considerable overlap in their imaging, endoscopic, and histopathological features [2, 3], the two diseases demand markedly different treatment strategies. In ITB-prevalent regions, patients are often empirically started on anti-tuberculosis therapy (ATT) and only switched to CD treatment if symptoms persist after about two months [4, 5]—a delay that can lead to clinical deterioration or immune-mediated side effects in misdiagnosed CD cases. Thus, early and accurate differentiation is of paramount importance.

While multiple diagnostic methods exist [6], fewer than half of patients exhibit the "classic" features of either disease [7]. Cross-sectional imaging modalities such as computed tomography enterography (CTE) provide a non-invasive means to visualize the entire intestine [8, 9] and overcome the limitations of endoscopy (e.g., accessing bowel segments distal to strictures). However, manual interpretation of CTE images—requiring the evaluation of markers like bowel wall thickness, comb sign, and calcified lymph nodes—is both time-consuming and subject to human error, especially in busy clinical settings.

Recent advances in deep learning and machine learning (ML) have shown promise in automatically extracting subtle imaging features that correlate with disease outcomes [10–15]. Yet, many existing approaches rely on manually delineated regions or incorporate additional clinical data, limiting their scalability and real-world applicability.

In this work, we introduce a cost-effective, radiology-only framework that leverages explainable ML to differentiate CD from ITB using non-invasive CTE scans. Our method employs a pretrained 3D segmentation model (TotalSegmentator) to automatically isolate regions of interest, followed by diagnostic modules that extract key radiological biomarkers—including the visceral-to-subcutaneous fat ratio [16], comb sign [17], calcified/necrotic lymph nodes, and pulmonary TB indicators. Notably, some of these features were overlooked during manual evaluations by experienced radiologists. Our approach also integrates interactive visualizations to localize disease sites, enhancing both the transparency and interpretability of the diagnostic process. Furthermore, the modular design of our system enables effective performance even with limited data (5–10 samples), thus paving the way for the automated annotation of larger deep learning datasets.

We evaluate our method using data from two hospitals: Hospital A (CD = 36, ITB = 57) for training and Hospital B (CD = 37, ITB = 28) for testing. Our feature-based approach is benchmarked against state-of-the-art (SOTA) models such as Google's CT Foundation model and a ResNet3D architecture. In addition, we employ XGBoost with SHAP to quantify the contribution of each feature.

Our main contributions are summarized as follows:
1. **Biomarker Detection:** We automatically detect critical radiological markers—namely the visceral/subcutaneous fat ratio, comb sign, calcified and necrotic lymph nodes, and a pulmonary TB indicator—to distinguish CD from ITB. Our technique successfully identifies subtle features that were missed by a team of radiologists during manual evaluation.
2. **Comparative Analysis of ML Algorithms:** We conduct a comprehensive comparison of diverse ML algorithms on the extracted features and employ XGBoost with SHAP to elucidate the contribution of each biomarker to the final prediction.
3. **Benchmarking Against SOTA:** We benchmark our approach against leading methods in intestinal disease detection, including a pretrained ResNet10 model (augmented with GradCAM for explainability) and CT-Foundation (finetuned Video-CoCa), demonstrating superior performance.

Table 1 summarizes key radiological features that have been used to distinguish CD from ITB in the literature [18]

**Table 1**: Radiological Features in Crohn's Disease vs. Intestinal Tuberculosis

| Feature | Parameter | Crohn's Disease | Intestinal Tuberculosis | Normal |
| --- | --- | --- | --- | --- |
| **Bowel Wall** | **Thickness** | 10–20 mm | 4–6 mm | up to 4 mm |
|  | **Symmetry** | symmetric | asymmetric | symmetric |
|  | **Mural Stratification** | yes | yes | no |
|  | **Mucosal Enhancement** | intense/irregular | mild | none |
| **Comb's Sign** |  | yes | no | no |
| **Lymph Node** | **Size** | 3–8 mm | >10 mm (peri-pancreatic / para-aortic) | – |
|  | **Center** | uniform | low-attenuation (necrosis or calcification) | uniform |
| **Associated Features** | **Peritonitis** | no | no | yes |
|  | **Splenic Involvement** | no | yes | no |

*Source:* Adapted from Haga et. al. [18]

Previous studies have predominantly relied on the manual interpretation of CTE images by radiologists [19–23], a process that is both labor-intensive and error-prone. Although earlier ML and deep learning approaches [10–15, 24] have improved diagnostic accuracy using handcrafted features, they typically depend on manual region identification or supplementary clinical data. To our knowledge, this is the first study to integrate fully automated, non-invasive CT imaging with independent diagnostic modules for the differentiation of CD and ITB.

In summary, our proposed framework not only enhances diagnostic accuracy and reduces radiologists' workload by automatically highlighting subtle imaging markers, but it also provides an explainable and robust methodology that can be extended to other intestinal disorders.

## 2 Methods

We begin by presenting the theoretical foundations and algorithmic justifications behind detecting biomarkers, detailing how the radiological features are detected, categorized, and leveraged for differential diagnosis. Then we propose multiple ML-based classifiers. To understand the contribution of each feature towards disease diagnosis, we also compute XGBoost and weight the contribution of each feature using SHAP.

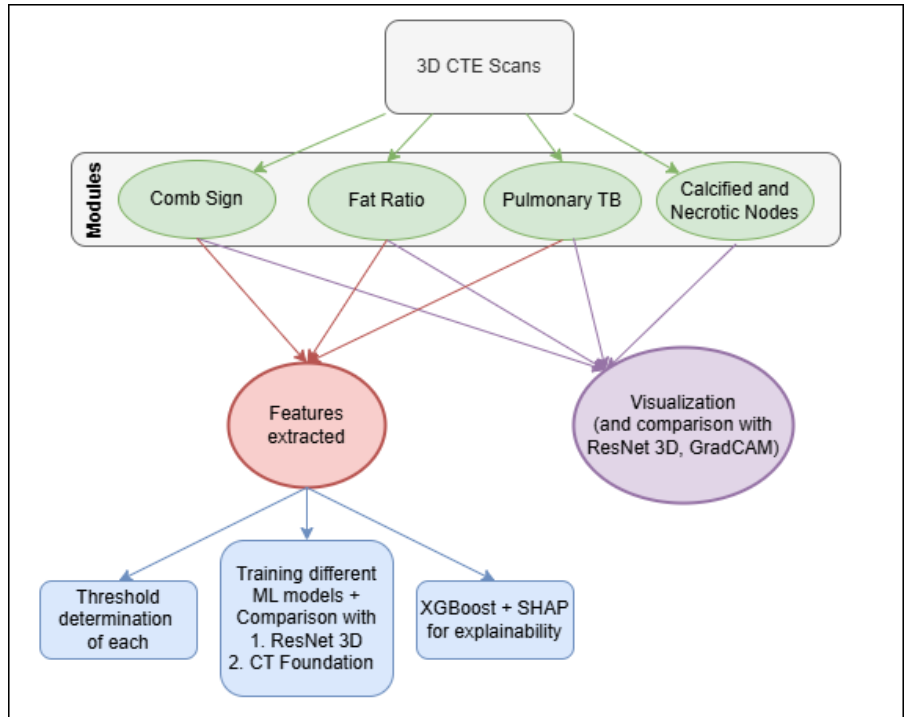

**Fig. 1**: Proposed Flow for CD vs ITB diagnosis

### 2.1 Detection of Biomarkers

Table 1, adapted from the Haga [18**?** ], summarizes the key radiological features essential for differentiating Crohn's disease from intestinal tuberculosis. In this section, we describe the modules developed to automate the detection of these features.

#### 2.1.1 Comb's Sign Detection

The Comb's sign is a characteristic radiological feature of Crohn's disease, reflecting localized hyperenhancement of the intestinal wall together with adjacent vasculature.

We define the voxel-wise probability of observing Comb's sign as

$$P(\text{voxel} \in \text{comb-sign}) = P(\text{voxel} \in \text{vessels}) \cdot P(\text{voxel is near enhanced wall}).$$

First, the intestine is segmented using TotalSegmentator. The intestinal wall is approximated by modeling the intensity histogram of the segmented region with a Gaussian Mixture Model (GMM). The optimal number of components is determined by minimizing the Bayesian Information Criterion (BIC)

$$\text{BIC}(k) = -2 \ln L_k + k \ln N,$$

where $L_k$ is the maximized likelihood for a model with $k$ components and $N$ is the number of data points. We empirically found that almost always the 4 gaussians include 2 high area modes for intestinal wall and intestinal contents, one for adjoining fat and a very distributed and low weight mode for other noise with very high standard deviation.

Vascular structures are detected via a vesselness filter based on Hessian analysis. For an image $I(\mathbf{x})$ at scale $s$, the Hessian is computed as

$$H_{ij}(\mathbf{x}, s) = s^2 \, I(\mathbf{x}) * \frac{\partial^2}{\partial x_i \partial x_j} G(\mathbf{x}, s),$$

with $G(\mathbf{x}, s)$ being the Gaussian kernel. The multiscale vessel response is then defined by

$$F(\mathbf{x}) = \max_{s \in [s_{min}, s_{max}]} V\left(\text{eig}(H(\mathbf{x}, s))\right),$$

where $V$ is an enhancement function favoring tubular structures.

To refine the vessel probability map $P^{(0)}(\mathbf{x})$, we perform an iterative update. Let the local maximum over a $3 \times 3 \times 3$ neighborhood be

$$M^{(k)}(\mathbf{x}) = \max_{\mathbf{y} \in N_3(\mathbf{x})} P^{(k)}(\mathbf{y}).$$

Then, the updated probability is computed as

$$\hat{P}^{(k+1)}(\mathbf{x}) = M^{(k)}(\mathbf{x})^{1-\lambda^{(k)}} P^{(k)}(\mathbf{x})^{\lambda^{(k)}},$$

with $\lambda^{(k)}$ balancing the contribution of the local maximum and the current value. After thresholding (typically at the 5% percentile), the final vessel map is modulated by a distance-dependent weight. This is achieved by convolving with a Gaussian kernel

$$f(r) = \sqrt{\frac{1}{2\pi\sigma^2}} \exp\left(-\frac{r^2}{2\sigma^2}\right),$$

where $r$ denotes the distance from the intestinal wall followed by setting the intestinal region as black since we don't want to detect tubular structures within intestine (eg. duodenum wall)

The resulting probability map highlights regions with high vascularity in close proximity to enhanced walls, thus serving as an effective indicator of the Comb's sign.

### 2.1.2 Fat-Ratio Estimation

We use the efficient method [25] to compute the visceral-to-subcutaneous fat ratio—a surrogate biomarker for differentiating Crohn's disease (CD) from intestinal tuberculosis (iTB). CT slices from the L4–L5 region are first thresholded between −500 and −50 Hounsfield Units (HU) to generate a binary fat mask. After applying two cycles of erosion and dilation to remove scanner artifacts, we determine the total fat area, $A_{total}$, and the subcutaneous fat area, $A_{subcut}$, via a polar boundary detection approach based on Bresenham's algorithm. In polar coordinates, the subcutaneous area is approximated by integrating along rays from the image center; for each angle $\vartheta$ (sampled with granularity $g$), the boundary is located by detecting the transition from fat to non-fat, and the local area is estimated as

$$\Delta A(\vartheta) \approx \frac{1}{2}\left[d_{out}^2(\vartheta) - d_{in}^2(\vartheta)\right]\Delta\vartheta,$$

where $d_{out}(\vartheta)$ and $d_{in}(\vartheta)$ denote the distances from the center to the outer and inner boundaries, respectively. Summing over all angles yields $A_{subcut}$.

The fat ratio is then computed as

$$\text{Fat Ratio} = \frac{A_{total}}{A_{subcut}} - 1.$$

Our algorithm runs in $O\left(\frac{V}{s \cdot g}\right)$ time, where $V$ is the image volume and $s$ is the number of axial slices, making it substantially more efficient and less artifact-prone than deep learning–based methods.

For disease classification, we optimize the threshold using Youden's J statistic:

$$J = \text{Sensitivity} - (1 - \text{Specificity}),$$

This volumetric approach, by aggregating ratios over multiple slices, improves robustness over single-slice analyses.

### 2.1.3 Pulmonary TB

Concurrent pulmonary and intestinal TB is rare [2]; thus, a positive pulmonary TB (PTB) finding strongly favors a diagnosis of intestinal TB (iTB). To detect PTB, chest slices are first isolated from 3D CT scans using a pretrained UNet-based TotalSegmentator [26]. These slices are then passed to the ViPTT-Net model [27]—a hybrid CNN-RNN architecture—whose final layer is replaced with a single-logit output. The probability of PTB is computed via a logistic function:

$$P_{PTB} = \frac{1}{1 + \exp(-z)},$$

where $z$ is the output logit. Despite the limited number of confirmed PTB cases (5 from Hospital A and 3 from Hospital B), the model yields high specificity (1.0000) but low recall (0.0303), indicating that a positive PTB detection nearly always shifts the diagnostic likelihood toward iTB.

### 2.1.4 Calcified Nodes

Calcified lymph nodes are detected through a series of image-processing steps. First, anatomical segmentation masks (in NIfTI format) are merged using a logical OR operation. A cubic structuring element is then applied to perform binary dilation on the merged mask, resulting in an expanded mask $M_{dilated}(x)$. The original CT image $I(x)$ is modified by setting voxels in the dilated region to a predefined Hounsfield Unit value $H_{calc}$:

$$I'(x) = \begin{cases} I(x), & x \notin M_{dilated}, \\ H_{calc}, & x \in M_{dilated}, \end{cases}$$

Afterward, thresholding at a calcification-specific level $T_{calc}$ yields candidate calcified voxels:

$$C(x) = \mathbf{1}\{I'(x) \geq T_{calc}\}.$$

Detection is restricted to abdominal slices with additional constraints to reduce edge artifacts.

### 2.1.5 Necrotic Nodes

Due to limited data, necrotic lymph nodes are approximated by detecting fluid-filled structures, which serve as a proxy for necrosis. Organ masks from TotalSegmentator are merged and dilated (excluding visceral fat) to define a region of interest (ROI). Within this ROI, voxels with CT intensities in the range $[T_{low}, T_{high}]$ are considered candidates for necrosis:

$$N(x) = \mathbf{1}\{T_{low} \leq I(x) \leq T_{high}\}.$$

Subsequent morphological operations—sequential erosion to remove noise followed by dilation to recover potential over-erosion—refine the candidate regions. Hyperparameters such as $T_{low}$, $T_{high}$, dilation width, and the number of erosion iterations are optimized to balance sensitivity and specificity.

## 2.2 ML Based Classifiers

### 2.2.1 Support Vector Machine (SVM)

Support Vector Machines aim to find an optimal hyperplane that separates two classes with maximum margin. The decision function is given by

$$f(x) = \text{sign}(w^\top x + b),$$

subject to

$$y_i(w^\top x_i + b) \geq 1, \quad \forall i.$$

The optimization problem can be formulated as

$$\min_{w,b} \frac{1}{2}\|w\|^2 + C \sum_{i=1} \xi_i,$$

where $C$ is a regularization parameter and $\xi_i$ are slack variables.

### 2.2.2 Logistic Regression

Logistic Regression models the probability of a binary outcome via the logistic function:

$$P(y = 1 \mid x) = \frac{1}{1 + e^{-(w^\top x + b)}}.$$

Parameters $w$ and $b$ are estimated by minimizing the cross-entropy loss,

$$L(w, b) = - \sum_{i=1} [y_i \log P(y_i \mid x_i) + (1 - y_i) \log (1 - P(y_i \mid x_i))].$$

### 2.2.3 Naive Bayes

Naive Bayes classifiers assume that features are conditionally independent given the class label. Using Bayes' theorem, the posterior probability is expressed as

$$P(C \mid x) \propto P(C) \prod_{j=1}^{d} P(x_j \mid C),$$

where $C$ denotes the class and $x = (x_1, \ldots, x_d)$ the feature vector. For continuous features, $P(x_j \mid C)$ is typically modeled using a Gaussian distribution.

### 2.2.4 Random Forest

Random Forest is an ensemble of decision trees where each tree is trained on a bootstrap sample with random feature selection. For classification, if $T_k(x)$ is the prediction of the $k$-th tree, the final prediction is given by majority voting:

$$f(x) = \text{mode}\{T_k(x)\}_{k=1}^{K}.$$

### 2.2.5 Gradient Boosting

Gradient Boosting builds an ensemble sequentially by adding weak learners that correct the errors of prior models. At iteration $t$, a new model $h_t(x)$ is fitted to the residuals, and the overall model is updated as

$$f_t(x) = f_{t-1}(x) + \eta\, h_t(x),$$

where $\eta$ is the learning rate. The method minimizes a differentiable loss function via gradient descent in function space.

## 2.3 Explainable Analysis

### 2.3.1 XGBoost

Consider a training dataset $\{(x_i, y_i)\}_{i=1}^n$ with features $x_i \in \mathbb{R}^d$ and labels $y_i \in \mathbb{R}$. XGBoost builds an ensemble of regression trees such that the prediction for instance $i$ is

$$\hat{y}_i = \sum_{t=1}^{T} f_t(x_i), \quad f_t \in \mathsf{F},$$

where $\mathsf{F}$ is the space of regression trees.

**Objective Function**

The overall objective to be minimized is

$$\mathcal{L} = \sum_{i=1}^{n} l(y_i, \hat{y}_i) + \sum_{t=1}^{T} \Omega(f_t),$$

with the regularization term defined as

$$\Omega(f) = \gamma T + \frac{1}{2}\lambda \|w\|^2,$$

where $\gamma$ and $\lambda$ are regularization parameters, $T$ is the number of leaves in the tree, and $w$ represents the vector of leaf weights.

At the $t$-th iteration, the prediction is updated as

$$\hat{y}_i^{(t)} = \hat{y}_i^{(t-1)} + f_t(x_i).$$

XGBoost employs a second-order Taylor expansion to approximate the loss for each instance:

$$l(y_i, \hat{y}_i^{(t-1)} + f_t(x_i)) \approx l(y_i, \hat{y}_i^{(t-1)}) + g_i f_t(x_i) + \frac{1}{2} h_i f_t(x_i)^2,$$

with

$$g_i = \frac{\partial l(y_i, \hat{y}_i^{(t-1)})}{\partial \hat{y}_i^{(t-1)}}, \quad h_i = \frac{\partial^2 l(y_i, \hat{y}_i^{(t-1)})}{\partial \hat{y}_i^{(t-1)2}}.$$

Assuming the new function can be written as

$$f_t(x) = w_{q(x)},$$

where $q : \mathbb{R}^d \to \{1, 2, \ldots, T\}$ maps inputs to leaf indices, let $I_j = \{i \mid q(x_i) = j\}$ be the set of instances in leaf $j$. Then, the optimal weight for leaf $j$ is

$$w_j^* = -\frac{\sum_{i \in I_j} g_i}{\sum_{i \in I_j} h_i + \lambda},$$

and the gain from splitting a leaf into left (*L*) and right (*R*) nodes is computed as

$$\text{Gain} = \frac{1}{2}\left[\frac{G_L^2}{H_L+\lambda} + \frac{G_R^2}{H_R+\lambda} - \frac{(G_L+G_R)^2}{H_L+H_R+\lambda}\right] - \gamma,$$

where

$$G_L = \sum_{i\in I_L} g_i, \quad H_L = \sum_{i\in I_L} h_i,$$

and similarly for $G_R$ and $H_R$.

### 2.3.2 SHAP (SHapley Additive exPlanations)

SHAP values provide a unified framework to interpret the output of any machine learning model by attributing each feature a contribution toward the final prediction. Any model $f(x)$ can be expressed as an additive explanation model:

$$f(x) = \phi_0 + \sum_{i=1}^{M} \phi_i,$$

where:
- $\phi_0$ is the base value (typically the expected model output, $E[f(x)]$),
- $\phi_i$ is the contribution of feature *i* (the SHAP value).

The SHAP value for feature *i* is given by the Shapley value formula from cooperative game theory:

$$\phi_i = \sum_{S\subseteq N\setminus\{i\}} \frac{|S|!(M-|S|-1)!}{M!}\left[f_{S\cup\{i\}}(x_{S\cup\{i\}}) - f_S(x_S)\right],$$

where:
- $N$ is the set of all features with $|N| = M$,
- $S$ denotes any subset of features not including *i*,
- $f_S(x_S)$ is the model output given only the features in $S$.

SHAP values satisfy several key properties:

(1) **Local Accuracy:** $f(x) = \phi_0 + \sum_{i=1}^{M}\phi_i$.
(2) **Missingness:** If a feature is absent or has no effect, its SHAP value is zero.
(3) **Consistency:** If a model changes so that the contribution of a feature increases (or does not decrease) for all subsets, its SHAP value will not decrease.

For tree-based models, the Tree SHAP algorithm leverages the structure of the ensemble to compute SHAP values in polynomial time, enabling scalable and interpretable explanations even for complex models.

## 2.4 Comparison with SOTA

### 2.4.1 Data Augmentation

To address the limited dataset size, we employ extensive data augmentation to synthetically expand the training samples. Our augmentation pipeline includes:

1. **Rigid Transformations:** We apply translations and rotations along the *x*, *y*, and *z* axes. A rotation is defined as

$$R(\alpha, \beta, \gamma) = R_z(\gamma) \, R_y(\beta) \, R_x(\alpha),$$

   where $R_x$, $R_y$, and $R_z$ denote the rotation matrices about the respective axes.

2. **Uniform Scaling:** Points are scaled uniformly about a center **c** as

$$\mathbf{x}' = \mathbf{c} + s\,(\mathbf{x} - \mathbf{c}),$$

   where *s* is the scale factor.

3. **Elastic Deformations:** Local warping is performed using a B-spline formulation:

$$\mathbf{u}(\mathbf{x}) = \sum_{i=0}^{n_x+3} \sum_{j=0}^{n_y+3} \sum_{k=0}^{n_z+3} \Phi_{i,j,k} \, B_i(x) \, B_j(y) \, B_k(z),$$

   with $B_i(\,)$ as the basis functions and $\Phi_{i,j,k}$ the control-point parameters.

4. **Intensity Augmentations:** To simulate variations in imaging conditions, Gaussian noise is added:

$$I'(\mathbf{x}) = I(\mathbf{x}) + \eta(\mathbf{x}), \quad \eta(\mathbf{x}) \sim \mathcal{N}(0, \sigma^2).$$

5. **Random Translation:** We randomly shift the coordinates along each axis by sampling offsets $\delta_x, \delta_y, \delta_z$ from a chosen distribution (e.g., uniform or normal). Each point **x** is thus transformed to

$$\mathbf{x}' = \mathbf{x} + (\delta_x, \delta_y, \delta_z).$$

By combining these transformations, we generate a diverse set of 3D samples, enhancing the model's ability to generalize across different anatomical and scanner variabilities.

### 2.4.2 ResNet10

ResNet10 is a lightweight variant of the Residual Network family designed for efficient 3D CT scan analysis. Its architecture is built upon residual blocks that learn the residual mapping:

$$y = F(x) + x,$$

where $F(x)$ represents the learned residual function. The skip (or shortcut) connections in ResNet10 alleviate the vanishing gradient problem, enabling effective training even with limited depth. Compared to deeper networks, ResNet10 offers lower computational overhead and faster inference, making it suitable for real-time applications and resource-constrained settings. Fine-tuning pretrained ResNet10 [28, 29] on our augmented dataset further improves its robustness and feature extraction capabilities. We use ResNet10 because of its SoTA performance in 3D imaging [30].

### 2.4.3 CTFoundation

CTFoundation is a state-of-the-art framework [31] that converts raw 3D CT volumes into compact, semantically rich embeddings. The process begins by reassembling individual 2D DICOM slices into a coherent 3D volume. It then employs an extended VideoCoCa model—a variant of the Contrastive Captioner (CoCa) that integrates both contrastive and captioning losses—to generate 1,408-dimensional embeddings. These embeddings encapsulate detailed spatial and semantic information about organs, tissues, and potential abnormalities, which greatly facilitates downstream classification tasks. The CTFoundation approach not only reduces computational overhead by providing efficient representations but also improves diagnostic performance, especially when labeled data are scarce.

**Table 2**: Threshold determination for comb-sign using different measurements along with metrics. We can see that the full-volume ratio over L4-L5 region gives the best results.

| Metrics | AUC | Threshold | Specificity | Sensitivity (Recall) | MCC | Accuracy | Balanced Accuracy |
|---|---|---|---|---|---|---|---|
| left sum | 0.6329 | 1833.6936 | 0.4828 | 0.7419 | 0.2181 | 0.5730 | 0.6123 |
| left ratio | 0.5651 | 0.0000 | 0.7241 | 0.4516 | 0.1771 | 0.6292 | 0.5879 |
| right sum | 0.5328 | 1268.0857 | 0.6207 | 0.5484 | 0.1624 | 0.5955 | 0.5845 |
| center sum | 0.5489 | 5505.4462 | 0.3793 | 0.7419 | 0.1222 | 0.5056 | 0.5606 |
| center ratio | 0.4627 | 0.0000 | 0.2586 | 0.8387 | 0.1111 | 0.4607 | 0.5487 |
| right ratio | 0.4850 | 0.0000 | 0.5172 | 0.5161 | 0.0318 | 0.5169 | 0.5167 |

**Table 3**: Threshold determination for fat-ratio by optimizing J-Statistics for each feature. We can see that the full-volume ratio over L4-L5 region gives the best results in terms of AUC.

| Metrics | AUC | Threshold | Specificity | Sensitivity (Recall) | MCC | Accuracy | Balanced Accuracy |
|---|---|---|---|---|---|---|---|
| Ratio | 0.8012 | 0.2969 | 0.4783 | 0.8485 | 0.3558 | 0.6964 | 0.6634 |
| Min fat-ratio | 0.7915 | 0.1696 | 0.5217 | 0.7576 | 0.2868 | 0.6607 | 0.6397 |
| Max fat-ratio | 0.7742 | 0.6685 | 0.6957 | 0.5455 | 0.2386 | 0.6071 | 0.6206 |
| Subcutaneous Fat | 0.6789 | 2733628.2263 | 0.8261 | 0.1818 | 0.0102 | 0.4464 | 0.5040 |
| Total Area | 0.6595 | 37748736.0000 | 1.0000 | 0.0000 | 0.0000 | 0.4107 | 0.5000 |
| Total Fat | 0.7392 | 5995958.0000 | 0.9565 | 0.0000 | -0.1615 | 0.3929 | 0.4783 |

## 3 Results

### 3.0.1 Biomarkers

We present the classification metrics for each of the biomarkers : Comb-sign, fat-ratio and pulmonary-tb.

**Comb Sign** Following the radiologists' recommendations, we compute the total sum of probabilities over the ileocecal junction (L3-S1 left) as well as anterior-central region and L3-S1 right regions, since they are the most likely regions of comb-sign. We compute the J-Statistic from the ROC curve for different regions to determine the best metric. L3-S1 left sum (ileocecal junction) presents the highest diagnostic performance. Since the volumes were normalized, sums (rather than volume densities) give better results. The thresholds and corresponding classification outcomes are presented in the table 2.

**Table 4**: Performance metrics for different models trained on features detected by our modules.

| Models | Accuracy | Balanced Accuracy | Recall | Specificity | PPV | F1 Score | MCC | AUC |
|---|---|---|---|---|---|---|---|---|
| Logistic Regression | 0.7500 | 0.7418 | 0.7879 | 0.6957 | 0.7879 | 0.7879 | 0.4835 | 0.7602 |
| SVM | 0.7321 | 0.7134 | 0.8182 | 0.6087 | 0.7500 | 0.7826 | 0.4383 | 0.7049 |
| Naive Bayes | 0.7143 | 0.6983 | 0.7879 | 0.6087 | 0.7429 | 0.7647 | 0.4030 | 0.7431 |
| Random Forest | 0.7143 | 0.6917 | 0.8182 | 0.5652 | 0.7297 | 0.7714 | 0.3984 | 0.6989 |
| Gradient Boosting | 0.6071 | 0.6271 | 0.5152 | 0.7391 | 0.7391 | 0.6071 | 0.2543 | 0.6469 |

**Fat-ratio** We adopt a threshold-based methodology to distinguish between Crohn's Disease (CD) and Intestinal TB (iTB). One dataset (Hospital A) is used for training, and the other (Hospital B) for testing. For each feature, we compute the ROC curve on the training set and select the point that maximizes Youden's J statistic:

$$J = \text{Sensitivity} - (1 - \text{Specificity}).$$

Using the resulting threshold, we classify the test set and report the metrics. For the *fat-ratio*, the optimal threshold is 0.29. The classification metrics on the test set are: AUC = 80.12%, Specificity = 47.83%, Sensitivity = 84.85%, MCC = 0.3558, Accuracy = 69.64%, Balanced Accuracy = 66.34%. The complete statistics (including the threshold) are shown in the following table 3.

**PulmonaryTB** In our dataset, there are 5 confirmed PTB cases from Hospital A and 3 from Hospital B. The key classification metrics for PTB as a feature are: Specificity = 1.0000, Recall (for PTB) = 0.0303, PPV Class 1 = 1.0000, PPV Class 0 = 0.4182, MCC = 0.1126, Accuracy = 0.4286, Balanced Accuracy = 0.5152.

**Calcified and Necrotic Nodes:** We developed simple proxy detectors for calcified and necrotic nodes. Since these nodes are not localized precisely, we do not present numerical feature extraction for them. Instead, these features appear as part of our visualization pipeline.

### 3.0.2 Comparison of performance of different ML Models based on our features

We compare multiple ML models trained on the above-described features. Logistic Regression achieves the highest Balanced Accuracy (0.7418), while SVM and Random Forest both attain the highest Recall (0.8182).

### 3.0.3 Results from SHAP and XGBoost

We train XGBoost on the numerical features (without SMOTE upsampling). When tested on Hospital B, XGBoost yields: AUC = 0.6693, Specificity = 0.7391, Sensitivity (Recall) = 0.5455, PPV Class 1 = 0.7500, PPV Class 0 = 0.5312, MCC = 0.2829, Accuracy = 0.6250, Balanced Accuracy = 0.6423. The SHAP plots are shown in Figures 2a and 2b, and a comparison of standard ML models on these features is given in table 4.

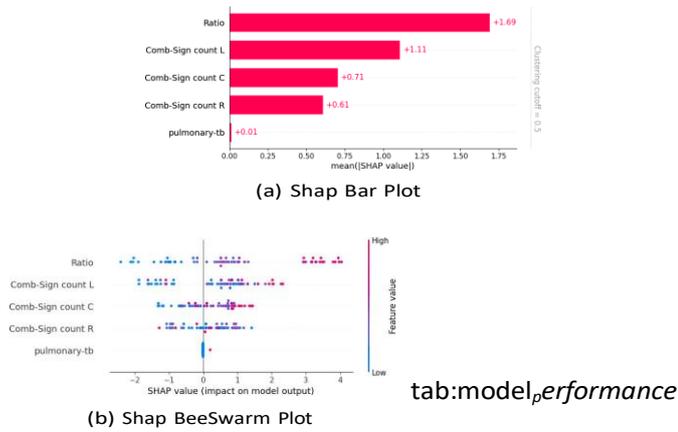

(a) Shap Bar Plot

(b) Shap BeeSwarm Plot

tab:model$_p$*erformance*

**Fig. 2**: SHAP plots showing the relative contribution of each feature.

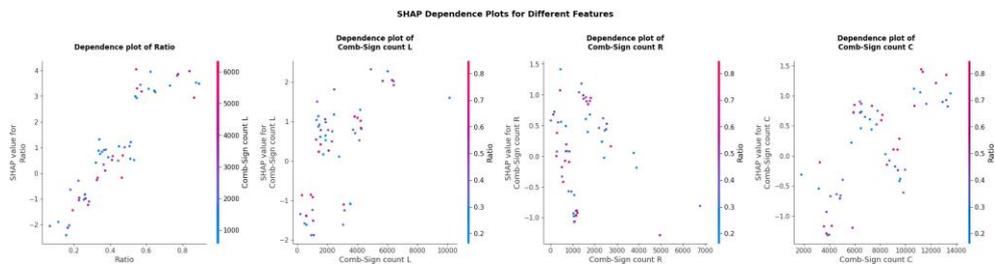

**Fig. 3**: Dependence plots of SHAP values with respect to feature variation.

### 3.0.4 Comparison with SOTA Models

We compare our Logistic Regression (trained on the proposed features) against two deep learning approaches, namely ResNet3D and CTFoundation (fine-tuned on our data). Table 5 summarizes the performance on the test set.

### 3.0.5 Rendering Visualiser

In the above sections, we compared our results with standard outcomes from SOTA models. However, we believe the full strength of our detected features is best appreciated in a 3D visualization context. We present snapshots of a 3D visualization tool that could assist doctors in clinical practice. From our team of three radiologists, the visualization tool detected certain features that were missed during manual examination. The visualization is shown in 7, with comb-sign displayed in red, fat-ratio slices in yellow, calcified nodes in green, and necrotic nodes in blue. Because we did not have calcified or necrotic nodes in this particular case, only comb-sign and fat-ratio are displayed. In contrast, GradCAM-based approaches (e.g., with ResNet) provide limited insights and do not clearly convey the spatial distribution of key features 8.

**Table 5**: Performance Comparison with other SoTA Models

| Model | Accuracy | Balanced Accuracy | Sensitivity (Recall) |
| --- | --- | --- | --- |
| Logistic Regression on features (ours) | 0.7500 | 0.7418 | 0.7879 |
| ResNet3D | 0.6163 | 0.5673 | 0.6250 |
| CTFoundation + FineTuning | 0.7404 | 0.7264 | 0.7406 |

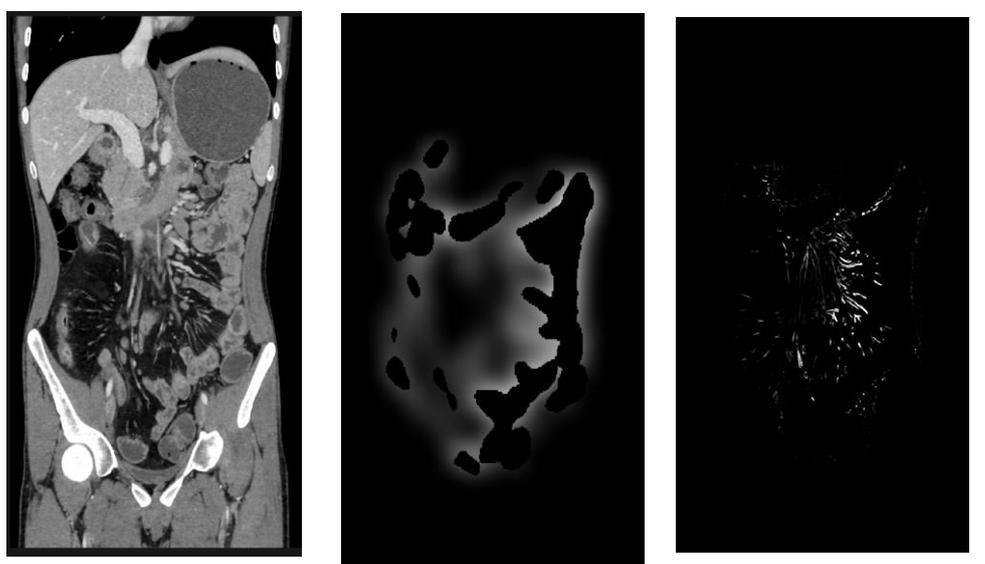

(a) Original Section showing comb-sign near enhanced bowel (mid-right of image)

(b) Probability plot of intestinal-wall enhancement with intestinal voxels set to zero, showing higher probabilities near the enhanced walls that decrease with distance.

(c) Efficient detection of Comb-Sign without involving other artifacts; higher probability means higher probability of comb-sign

**Fig. 4**: Result of detecting comb-sign. We can find that the module is able to detect only the hyper-vascularity close to enhanced-bowel

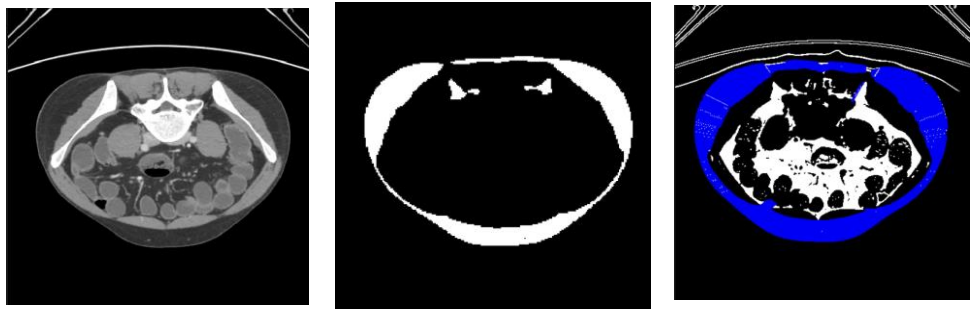

(a) Original CT Scan (Windowed for visibility).
(b) Segmentation of Subcutaneous fat by TotalSegmentator (with erroneous visceral fat)
(c) Segmenting the subcutaneous fat using our module (minimal error)

**Fig. 5**: Comparison of our module with the TotalSegmentator (DeepLearning) based methods for the same slice. Though known for generalisability, DeepLearning methods can at times introduce erroneous predictions such as we see here.

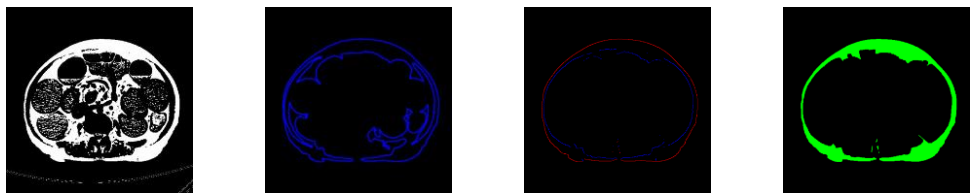

(a) Original CT Scan (binarized based on HU).
(b) Marking outer boundary (skin) based on contours (erroneous).
(c) Marking outer and inner boundaries using our algorithm.
(d) Segmenting the subcutaneous fat using our CV algorithm (minimal error).

**Fig. 6**: Comparison of our algorithm with the classical approach of contours, and segmentation of subcutaneous fat using our algorithm.

## 4 Discussion

### 4.0.1 Insights

As expected, the comb-sign probability sums in the left (ileocecal) region produce the most reliable metric, highlighting the disease location's importance. The shift in threshold and diagnostic performance across different regions likely reflects anatomical and pathological variations. For the fat-ratio, the new 3D volume-based threshold (0.29) outperforms the previously reported threshold of 0.63, enhancing sensitivity by about 4% points. This volume-based approach is also more straightforward to implement than searching for a "max-fat local segment" slice by slice.

The PTB feature shows high specificity but low recall, suggesting that the presence of confirmed pulmonary TB in our dataset is rare and not sufficient for robust discrimination of intestinal TB. Due to limited PTB samples, SHAP-based estimates for this feature should be interpreted cautiously.

Although SHAP is valuable for feature-level interpretability, its assumption of feature independence complicates spatially correlated phenomena (e.g., the comb-sign). Interpreting SHAP values for large 3D volumes can also be challenging for radiologists. Consequently, a visualization-based framework that highlights suspicious regions directly on the 3D scan may offer clearer interpretability.

Finally, comparing classical ML methods with SOTA deep learning approaches (e.g., ResNet3D, CTFoundation) shows that hand-crafted features capturing core radiological biomarkers (comb-sign, fat-ratio) can achieve competitive performance. This finding underscores the continued value of interpretable features in clinical decision-making.

### 4.0.2 Ethical and Societal Impacts

Automated detection and classification tools can expedite clinical workflows and reduce diagnostic ambiguity. However, ethical considerations such as patient privacy, data governance, and potential biases in algorithms trained on region-specific data must be addressed. In resource-limited settings, these methods could improve diagnostic efficiency but also risk exacerbating inequities if not carefully validated and deployed. Ongoing collaboration with clinicians, ethicists, and policymakers is essential to ensure responsible integration of AI-driven diagnostic aids.

### 4.0.3 Rendering Visualiser

While we compared our results with standard outcomes from SOTA models, we believe the true strength of our detected features is best appreciated in a 3D visualization context. We present snapshots of a 3D visualization tool that could assist doctors in clinical practice. From our team of three radiologists, the visualization tool detected certain features that were overlooked during manual examination. The visualization is shown in 7, with comb-sign in red, fat-ratio slices in yellow, calcified nodes in green, and necrotic nodes in blue. Because we did not have calcified or necrotic nodes in this particular case, only comb-sign and fat-ratio are displayed. In contrast, GradCAM-based approaches (e.g., with ResNet) provide limited insights into the spatial distribution of key features 8.

### 4.0.4 Limitations

Despite its promise, our study has several limitations. First, the dataset is relatively small and sourced from only two South Asian hospitals, which may limit the generalizability of our findings. Second, some modules rely on empirically set parameters that might not be optimal under varying imaging conditions or across diverse patient populations. Lastly, the proxy detectors for necrotic and calcified lymph nodes—while helpful for visualization—produce a significant number of false positives and need further refinement.

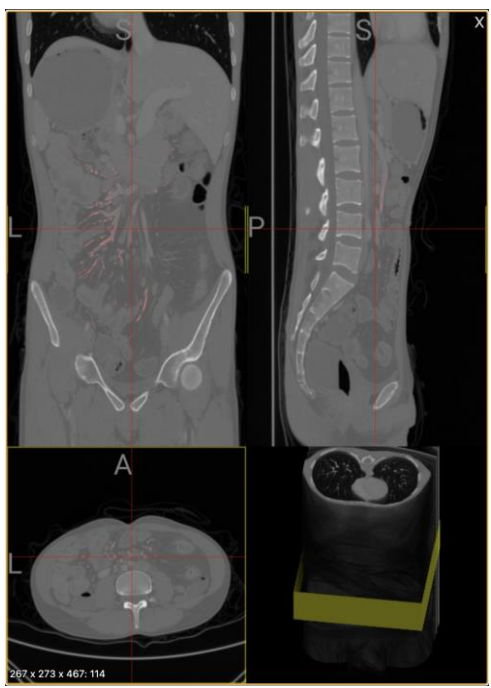

**Fig. 7**: Rendering Visualiser

### 4.0.5 Future Directions

Building on these initial results, future research may focus on:
1. **Algorithmic and Computational Enhancements:**
    - Optimize implementation by leveraging parallel processing and refactoring code to reduce runtime, which is crucial for processing multiple clinical cases simultaneously.
    - Investigate dynamic hyperparameter tuning methods (e.g., attention mechanisms) to automatically adapt parameters for each image.
2. **Expansion of Feature Modules:**
    - Incorporate additional radiological biomarkers to further discriminate between Crohn's Disease (CD) and Intestinal TB (ITB).
    - Validate and adapt the algorithms on datasets from geographically diverse and demographically varied populations.
3. **Advanced Deep Learning Integration:**
    - Integrate deep learning approaches that automatically learn optimal features, reducing reliance on empirically set parameters.
    - Explore hybrid models combining classical image processing, deep learning, and multimodal data (e.g., clinical history, lab values). Features with presets could also serve as pseudo-labels in an active learning framework.
4. **Clinical Translation and User-Centered Development:**

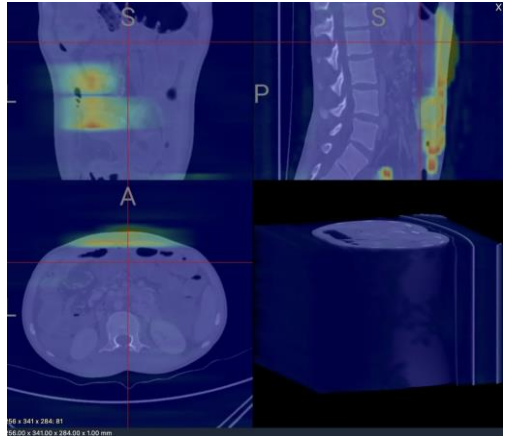

**Fig. 8**: GradCAM for ResNet

- Collaborate with HCI experts to design intuitive interfaces that facilitate the visualization and annotation process for clinicians.
- Conduct thorough clinical trials and user studies to validate the system's impact in real-world medical environments.

## 5 Conclusion

In this work, we introduced **EXACT-CT**, a novel and explainable framework for differentiating Crohn's Disease (CD) from Intestinal Tuberculosis (ITB) using CT Enterography images. Our approach combines classical image processing techniques with deep learning methods to automatically extract and analyze key radiological biomarkers—such as the visceral-to-subcutaneous fat ratio, Comb's sign, and pulmonary tuberculosis evidence—as well as proxy detectors for necrotic and calcified lymph nodes. By coupling these feature-detector modules with an interpretable classifier (XGBoost enhanced with SHAP analysis), EXACT-CT achieves competitive diagnostic performance compared to state-of-the-art 3D deep learning models, while also offering clear visual and numerical insights that support clinical decision-making. The accompanying 3D visualization tool further aids radiologists by highlighting subtle imaging features that can be missed in high-volume workflows. Overall, our results suggest that EXACT-CT can reduce diagnostic ambiguity, minimize reliance on empirical treatment trials, and ultimately improve patient outcomes.

**Author contributions statement.** Sh.G. did the experiments and came up with the modules. Sa.G. and M.N. helped with the understanding of radiological features and correlating features with their observable characteristics. A.A. assisted in research work, identifying relevant models and editing and presenting the manuscript. P.B. and R.Y. helped in advising on the algorithms as well as radiological features. Sa.G., M.N, and R.Y. are the radiologists consulted. All authors reviewed the manuscript.